# Multi-Channel Amplitude-Phase Asymmetric-Encrypted Janus Acoustic Meta-Holograms

Haohan Zeng[1], Zhenyu He[1], Tianxiang Zhang[2], Xiao Guo[1], Xinghao Hu[1], Youyu Mo[1], Tingting Li[1], Feilong Mao[1], Haiyan Fan[1], Xudong Fan[3], Weiwei Kan[4]*, Yifan Zhu[1]*, Hui Zhang[1]*, Guodong Yin[5], Badreddine Assouar[6]*

[1]*Jiangsu Key Laboratory for Design and Manufacturing of Precision Medicine Equipment, School of Mechanical Engineering, Southeast University, Nanjing 211189, China.*

[2]*School of Cyber Science and Engineering, Southeast University, Nanjing 211189, China.*

[3]*National Key Laboratory of Transient Physics, Nanjing University of Science and Technology, Nanjing 210094, China.*

[4]*School of Science, MIIT Key Laboratory of Semiconductor Microstructure and Quantum Sensing, Nanjing University of Science and Technology, Nanjing 210094, China.*

[5]*School of Mechanical Engineering, Southeast University, Nanjing 211189, China.*

[6]*Université de Lorraine, CNRS, Institut Jean Lamour, Nancy, 54000, France.*

**\*Corresponding authors:**

\*kan@njust.edu.cn

\*yifanzhu@seu.edu.cn

\*seuzhanghui@seu.edu.cn

\*badreddine.assouar@univ-lorraine.fr

**Abstract**

Encrypted optical and acoustic meta-holograms only focus on the encrypted hologram in a single channel, viz. modulating spatial amplitude to project a holographic image. In this research, the unique concept of multi-channel amplitude-phase asymmetric-encrypted Janus acoustic meta-holograms is proposed, demonstrating remarkable capabilities of generating, encrypting, and decrypting both amplitude and phase holographic images on both sides of a metascreen. The flexible and decoupled manipulation mechanism for the amplitude-phase of the bidirectional acoustic waves used in our concept offers multiple possibilities to apply various encryption methods. In this work, our system enables single-input, two-faced four-channel asymmetric encryption, which substantially increase the communication capacity of conventional acoustic holograms, and establish a security framework based on mathematical problem, proving its security. Our work can lead to concrete applications including, but not limited to, multi-channel acoustic field communications and acoustic illusion and cloaking in non-transparent media.

**Introduction**

The pursuit of acoustic wave precise manipulation has long captivated scientific inquiry. Metamaterials show a significant advantage in this area, which stems from their sub-wavelength artificial structure that allows the manipulation of the acoustic wavefront in an arbitrary manner[1-11]. Acoustic metamaterials can project pre-designed acoustic field distributions to predetermined locations to form acoustic holograms[12-17]. In the conventional acoustic holography, researchers only focus on the amplitude distributions and their corresponding physical effects. For instance, acoustic holography was developed as acoustic tweezers for manipulating suspended particles without contact[18,19], and acoustic holography has been used to manipulate bubbles in liquids to achieve dynamic holographic images[20].

However, if we consider the hologram as information for communication, the physical effects of the amplitude distribution are not important[21]. On the contrary, the phase is also a valuable information carrier whose distribution contains information as well as the amplitude. Therefore, we proposed the concept of amplitude-phase dual-channel encrypted acoustic meta-holograms in a previous work[22]. The encrypted communications capacity can be expanded by introducing phase channel, and further increasing its capacity is still a challenging problem. Janus metascreen[11, 23-25] for bidirectional modulation is a way to increase its channel number, although its conventional designs focus more on phase modulation while unable to control the amplitude very precisely.

In terms of encrypted holograms, optical meta-holograms have been widely investigated for the application in optical information confidential communication[26-39], while most of the studies of such holographic encryption lack rigorous mathematical model derivations to prove their high security. In the specific environments with strong light and electromagnetic interferences, acoustic wave becomes good carriers of information in these media. Researchers have also used optical and other methods to construct encryption systems for encrypting acoustic information[40-43]. Comparing with above methods, the acoustic-metamaterial-based asymmetric encryption holograms, although costing more in material fabrication, have several advantages, such as extremely high channel capacity expansion, higher security than symmetric encryption systems, and more security protocol expansion.

In this work, we report on a framework for encrypted acoustic meta-holograms with the advantages of high communication capacity, flexible encryption methods, and strong robustness and security, by introducing acoustic-metamaterial-based asymmetric encryption holograms[22] with Janus[11] and simultaneous amplitude and phase[14] modulations. For a conventional acoustic meta-hologram system, it tends to be a single-input single-output system, i.e., one metamaterial corresponds to one hologram output. As shown in Fig. 1, our designed multi-channel Janus encrypted acoustic hologram (JEAM) as a single-input multiple-output system can support up to four channels of information encryption and decryption in both directions (transmitted and reflected directions). In this system, we adopt the asymmetric encryption by using JEAM as the public key to achieve four channel encryptions of amplitude-phase in both directions, and the corresponding four metakeys as the private key. JEAM enables to decouple amplitude-phase modulation of transmitted and reflected waves, and each amplitude or phase can be controlled by a unique structural parameter[44,45]. It means that the amplitude-phase distribution of the reflected and transmitted sound fields can be designed arbitrarily, which provides extreme flexibility for encryption methods[15]. On the other hand, these private metakeys are used to decipher secure information and recover predesigned images in corresponding channels with high fidelity. For the convenience of the experimental demonstrations and sample fabrications, we choose the working frequency of 40 kHz at ultrasonic regime[46].

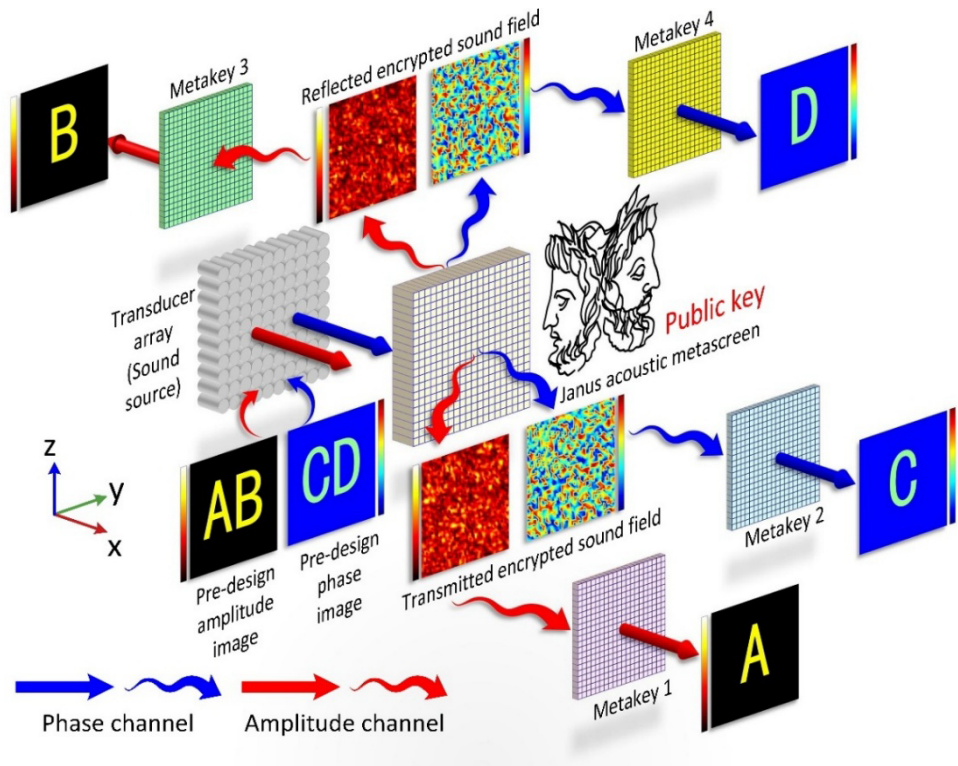

**Fig.1 Concept of Multi-Channel Amplitude-Phase Asymmetric-Encrypted Janus Acoustic Meta-Holograms.** Schematic diagram of the multi-channel amplitude-phase asymmetric encrypted Janus acoustic meta-holograms. The blue and red arrows indicate the Phase and amplitude channel. The winding arrows indicate the channels of encrypted sound field.

## Results

**Coupling effects between amplitude and phase channels.**

Acoustic hologram system includes the elements of pre-deign images, hologram plane, and the image plane, where the image plane and the pre-designed image comprise pixels, and the hologram plane may be acoustic metascreen or acoustic emission arrays. Therefore, the physical quantities of all three elements in the acoustic hologram system can be presented as matrices. Firstly, the predesigned images have an amplitude and phase distributions at image plane, defined as ($A^i_{x,y}$, $\varphi^i_{x,y}$), where $x$ and $y$ are the line and row numbers of each image pixels. When the size of the metascreen cell's cross section is subwavelength, each pixel in the image plane can be considered as a superposition of spherical waves emitted by the metascreen cells. Accordingly, each cell of the metascreen is a point source of a spherical wave, and the spherical wave function can be expressed as

$$p = \frac{A^h_{m,n}}{r_{m,n}} \exp\left(ikr_{m,n} - i\varphi^h_{m,n}\right) \quad (1)$$

Where ($A^h_{m,n}$, $\varphi^h_{m,n}$) is amplitude and phase distributions at the hologram plane, $m$ and $n$ are the row and column numbers of each cell on the hologram plane. Furthermore, the spherical wave superposition at a point in the half-space far field can be expressed as (Supplementary Note 1 for derivation and boundary conditions of the Equation (2).)

$$p_{x,y,z} \equiv A_{x,y,z} \exp\left[i\varphi_{x,y,z}\right] = \sum_{m=1}^{m}\left[\sum_{n=1}^{n} \frac{A^h_{m,n}}{r_{m,n}} \exp\left(ikr_{m,n} - i\varphi^h_{m,n}\right)\right] \quad (2)$$

Where ($x$, $y$, $z$) are the coordinates of a point in the modulated sound field, and $r_{m,n}$ is the spatial distance between this point and the cell on the hologram plane.

$$r_{m,n} = \sqrt{\left(x - x_{m,n}\right)^2 + \left(y - y_{m,n}\right)^2 + \left(z - z_{m,n}\right)^2} \quad (3)$$

Then, as shown in Fig. 2, we can make ($A_{x,y}^i$, $\varphi_{x,y}^i$) and ($A_{m,n}^h$, $\varphi_{m,n}^h$) in the same spatial coordinate system. According to the grayscale matrix of the pre-design images, the time-reversal method is used to obtain the required distribution ($A_{m,n}^h$, $\varphi_{m,n}^h$) at hologram plane.

Since the amplitude and phase of the pre-designed images are independent, they can be treated as two separate channels. The ideal situation is shown in Fig. 2a showing that two independent channels can project images without interfering with each other. However, as shown in Fig. 2b, when the pre-designed amplitude and phase of the holograms both contain image information, the information transmitted by the two channels will be coupled. This means that the correct information is not available to the receivers of both channels and can result in information leakage, which is a fundamental issue in confidential communications. Indeed, when there is a pre-designed image in the amplitude channel, it is inevitable that there will be a similar image in the phase channel, as has been demonstrated in many previous acoustic holography studies[8, 12]. So in the phase distribution ($\varphi_{x,y}^i$) of Fig. 2b, the letters AB and the continuous phases around them can be clearly observed. This suggests that interference from one channel to another is inevitable when transmitting acoustic holograms normally.

Thus, there is a mutual interference and coupling between channels, which also leads to the leakage of information transmitted in the channel.

In the following, we will design a Janus encrypted acoustic holograms system that not only solves the channel coupling problem also increases the communication capacity.

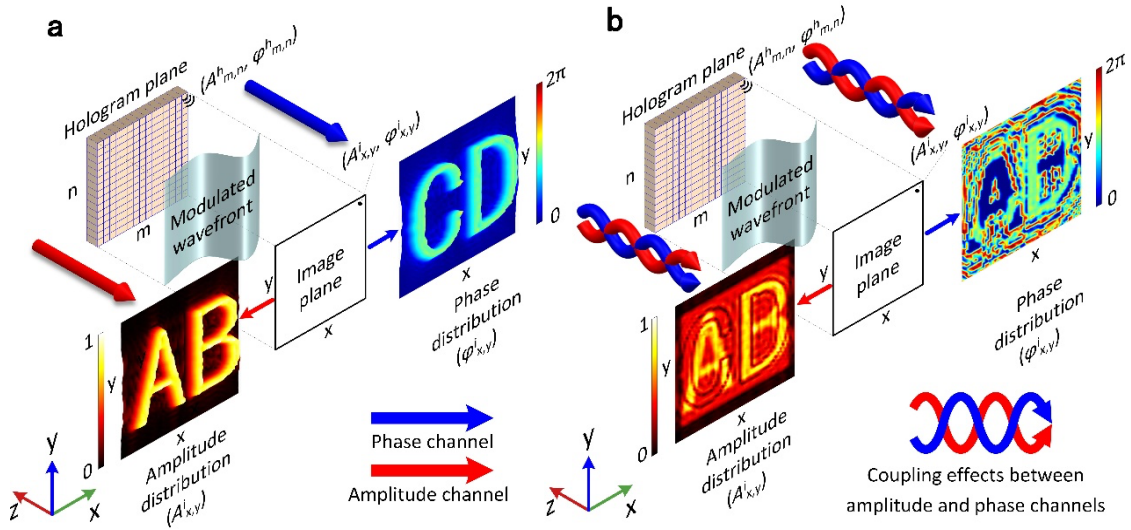

**Fig.2 Dual-channel acoustic hologram imaging process. a** In the ideal case the amplitude and phase channels are independent of each other and transmit separate images. **b** In the real situation the amplitude and phase channels interfere with each other, and coupled images appear in both channels causing information leakage.

**Deign of Janus encrypted acoustic hologram system.**

Figure 3 shows the encryption and decryption processes of this system. The transmitted field (original information) created by the transducer array (hologram plane) is unencrypted and coupled, and the receivers can obtain the hologram information directly. The hologram containing all readable information of the orignal transmitted field will be then enciphered into two encoded holograms based on the flexible encryption methods by the Janus encrypted metascreen. (Supplementary Note 2 for details of the encryption and decryption processes and the mathematical framework.) As a result, the encrypted channels contain unreadable holograms characterized by the projected acoustic field. JEAM plays the role of public key here, and we just need to place the corresponding private key (Metakey) in the encrypted field to get the decrypted image, which is exclusive, clean and decoupled.

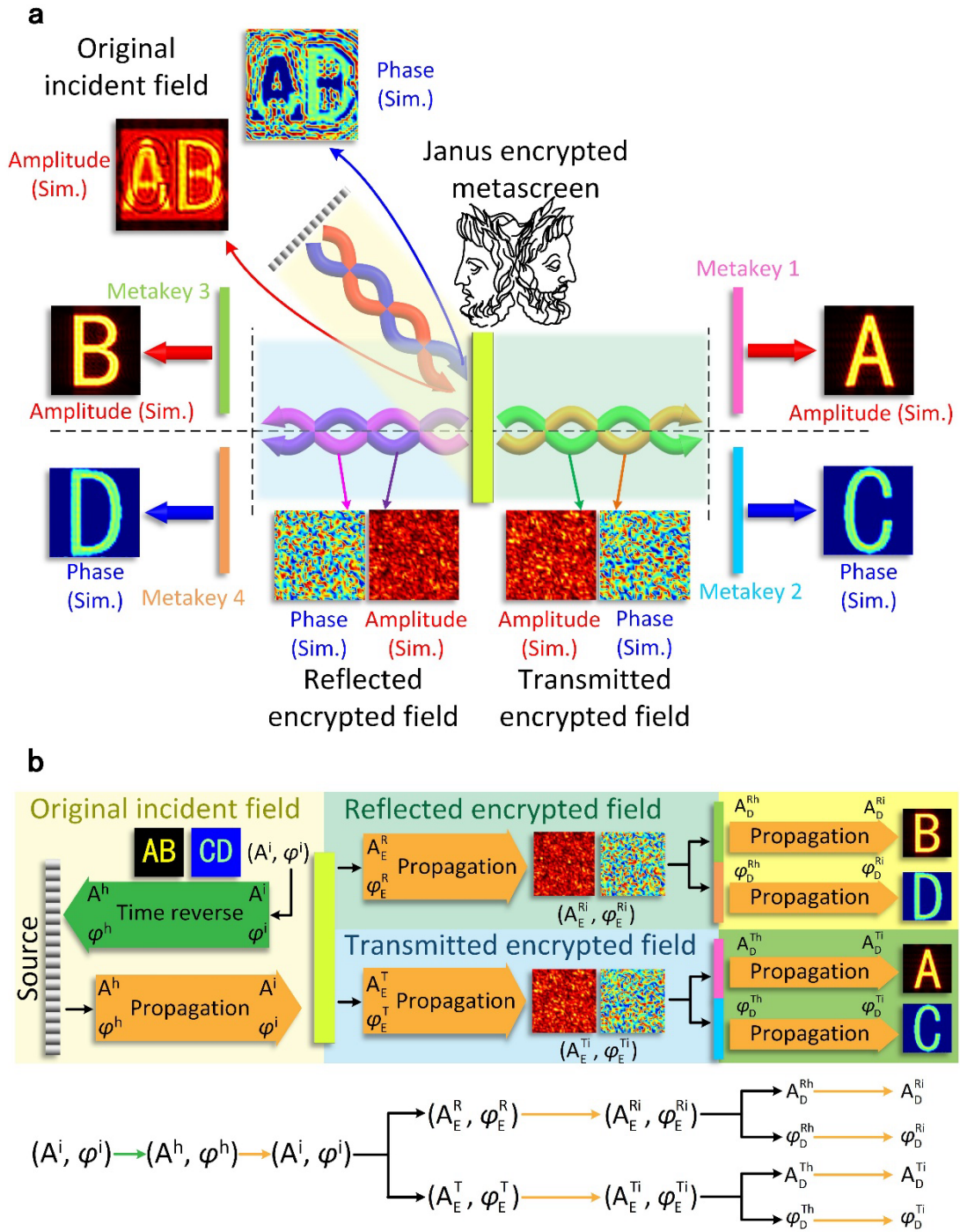

**Fig.3 The encryption and decryption processes of Janus encrypted acoustic hologram system. a** The distribution of the sound field in the system and the numerical simulation results of the holograms. **b** The encryption and decryption flowchart of the system, where the superscript *i* represents the image plane, *h* represents the holographic plane, *R*

represents reflection, *T* represents transmission, and the subscript *E* represents encryption and *D* represents decryption.

Obviously, JEAM is the core of the whole system. It consists of *M*×*N* unit cells, which means its inputs and outputs can be indicated by matrices with *M*×*N* elements. In the original transmission field coordinate system, the amplitude and phase received by the cell of JEAM can be described as

$$A^i_{x,y,z} = A^I_{M,N} \tag{4}$$

$$\varphi^i_{x,y,z} = \varphi^I_{M,N} \tag{5}$$

Where $A^i_{x,y,z}$ and $\varphi^i_{x,y,z}$ can be obtained by Eqs. (2), *M* and *N* are the row and column numbers of each cell on the JEAM.

The details for the metascreen designs are shown below. The unit cell shown in Fig. 4a-b can be divided into two parts, the reflected wave modulation part and the transmitted wave modulation part. In Fig. 4c-h, Each part is able to independently modulate the amplitude and phase of the corresponding acoustic wave, which means that $A^R_{M,N}$, $\varphi^R_{M,N}$ and $\varphi^R_{M,N}$ of the encrypted acoustic wave are related to the unit cell structure parameters *w*, *p* and *a* (Supplementary Note 3 for theoretical model). From the results, the $A^R_{M,N}$ can cover the span from 0 to 1 when varying the structure parameter *w* while keeping the $\varphi^R_{M,N}$ unchanged. Similarly, the $A^R_{M,N}$ does not change when varying the structural parameter *p*, while the span of the $\varphi^R_{M,N}$ can be covered by 2π. The $\varphi^T_{M,N}$ also controlled by the *a* from 0 to $2\pi$. $A_{M,N}^T$, on the other hand, can be calculated from the following equation based on the conservation of energy.

$$A_{M,N}^T = 1 - A_{M,N}^R - A_{M,N}^L \left(1 - A_{M,N}^R\right) \tag{6}$$

Where $A_{M,N}^L$ means the loss coefficient of amplitude, and it is controlled by the absorbing sponge, which has been demonstrated to be considered equivalent to a resistance but not as a reactance to acoustic waves when the thickness is subwavelength, and therefore has nearly no effect on the phase[47-49]. The five parameters Johnson-Champoux-Allard (JCA) model employ are homogeneous material parameters to effectively characterize the behavior of sound-wave propagation within the porous medium, which is provided in the Supplementary Note 4. Figure 4 does not only indicate the decoupling feature but also the moderating ability of the absorbing sponge. It is worth stating that the $A_{M,N}^T$ is obtained at a certain value of $w$ and taking into account the space constraints, thus, the moderating capacity of the absorbing sponge is limited. Figure 4c shows a matrix composed of the amplitudes received by each unit cell of the JEAM. Since the amplitude of sound waves is positively correlated with acoustic energy, amplitude is used here to represent the manipulation and distribution of energy. (Supplementary Note 5 for the specific examples of acoustic energy manipulation and distribution.) In this way, we achieve decoupled modulation of the encrypted acoustic waves. This extremely flexible manipulation of bidirectional acoustic waves gives us a very high degree of freedom in the design of the encryption methods.

The periods of unit cells is $d = 0.25\ \lambda$ for the chosen operating wave. Because *d* is less than half-wavelength, the local reflected amplitude and phase are independent of the incidence direction, which means the angle between the incident and reflected directions on JEAM is changeable. It is worth noting that interactions between unit cells can affect the quality of holograms, which is inevitable in most cases. In our design, we have extensively suppressed the interaction through reasonable structural and dimensional design. The detailed solutions and derivations are provided in Supplementary Note 6.

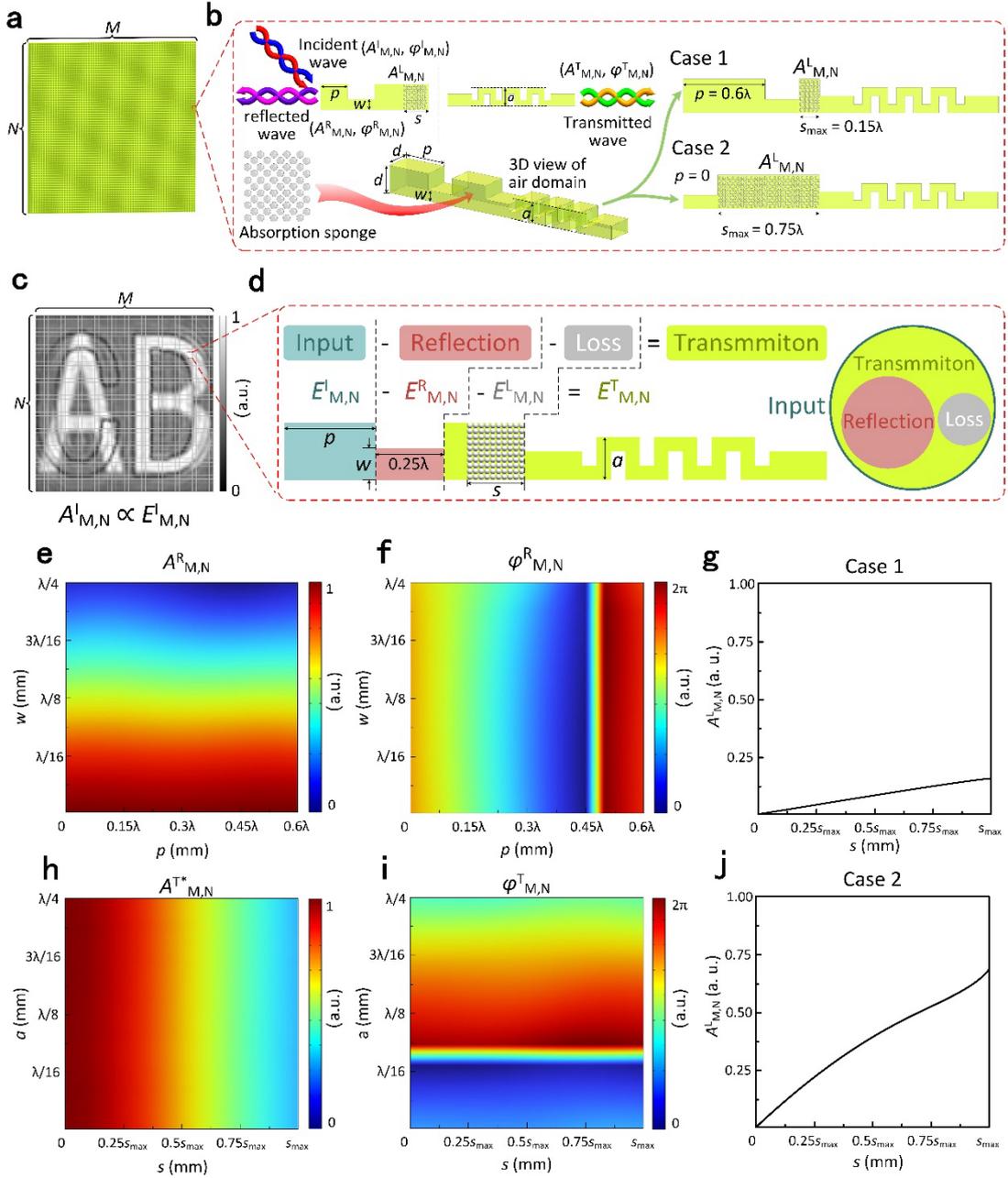

**Fig.4 Bidirectional acoustic wave amplitude-phase modulation encryption for JEAM. a** Geometric schematic of JEAM. **b** Details of JEAM unit cell. The reflected wave phase $\varphi_{M,N}^{R}$ is controlled by $p$, the reflected wave amplitude $A_{M,N}^{R}$ is controlled by $w$, the transmitted wave phase $\varphi_{M,N}^{T}$ is controlled by $a$, and the transmitted wave amplitude $A_{M,N}^{T}$ is controlled by the $s$ (thickness of absorption sponge) according to the law

of conservation of energy **c** The matrix received by JEAM from the amplitude channel of the transducer array. **d** Distribution and manipulation of acoustic energy in JEAM unit cell. The matrix received by JEAM from the amplitude channel of the transducer array. **e-j** These simulation results reveals that the amplitude and phase are controlled by only one parameter, respectively. $A_{M,N}^{T*}$ is the ratio of the transmitted wave amplitude to the amplitude before loss. Figures are carried out by the finite element solver in commercial software COMSOL Multiphysics 6.2. Case 1 and Case 2 demonstrated the ability of acoustic sponges to modulate losses in two extreme situations.

Similar to the JEAM, the metakeys are composed of a series of 3D unit cells made of 3D printed metasurface. For the different channels, there are different metakeys. These metakeys can also separately modulate the amplitude and phase in the encrypted field, which can directly reconstruct the decoded acoustic field. As shown in Fig. 5, taking the transmitted sound field of JEAM as an example, the random matrices are used to construct the encryption matrices to encrypt the sound field information, and then reconstruct the sound field through the corresponding metakeys to obtain the corresponding image information. The decryption process for the reflected encrypted sound field is similar to Fig. 5 and will be shown in the Supplementary Note 7.

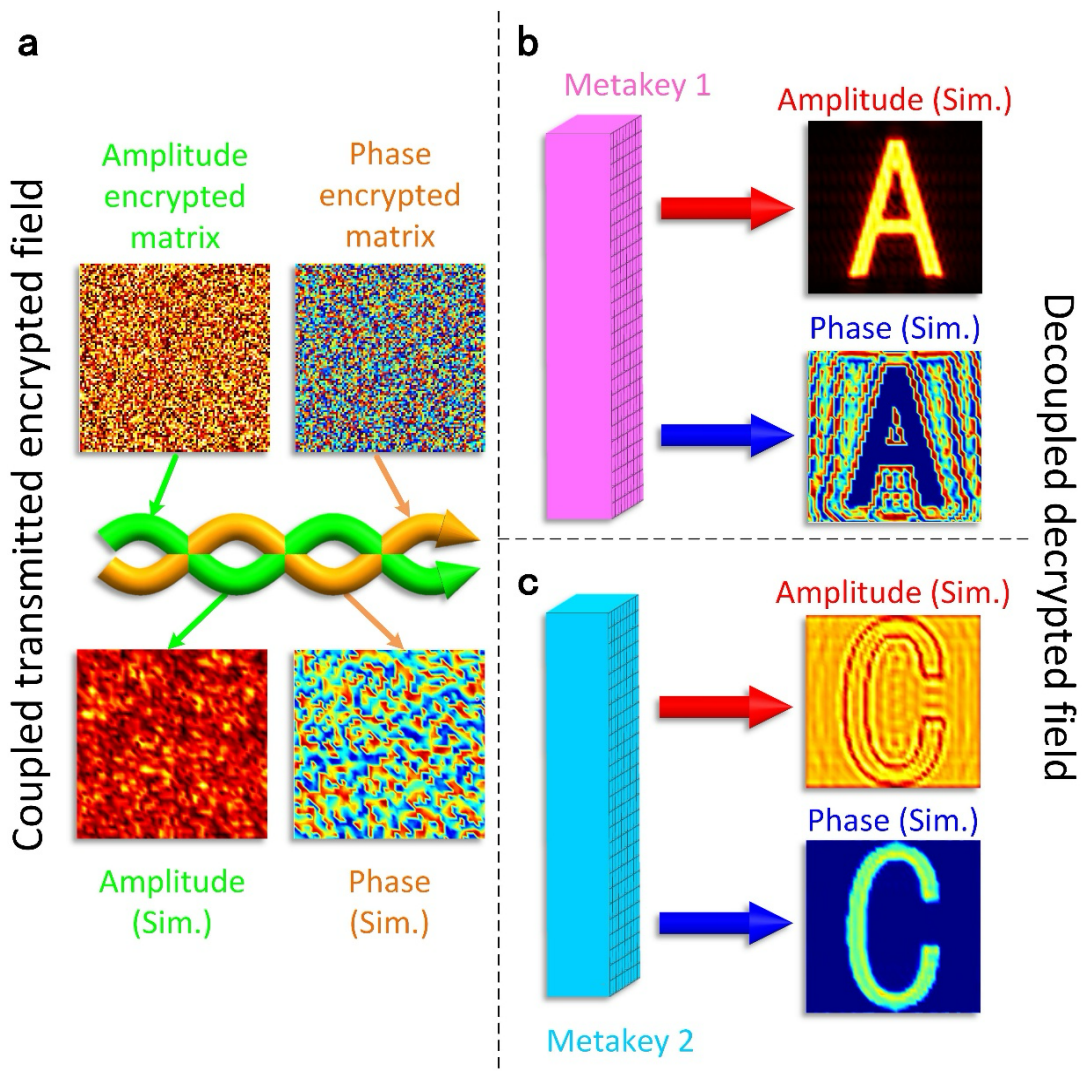

**Fig.5 Metakey 1 and metakey 2 reconstructs the transmitted decrypted acoustic field. a** We obtained the encrypted sound field by the random matrices, where both the amplitude and phase distributions (channels) are unreadable images. **b** Metakey 1 decrypted the letter A, and the phase distribution also indicates only the letter A. **c** Metakey 2 decrypted the letter C, and the amplitude distribution also indicates only the letter C.

Interestingly, the metakeys can decouple the two mutually coupled channels observed in Fig. 2b in the encrypted transmitted sound field, and decode the encrypted information in the corresponding channels successfully without any undesired coupling effects. It is worth noting that the metakey 1 as an amplitude key, which decrypts the amplitude information in the transmitted encrypted sound field, i.e., the letter A. The letter A can also be seen in the phase distribution in the decrypted sound field, which is determined by the nature of the sound field, which has been proved by many researches. However, this is not the information we are interested in, and it is important that we find no interference in it, which means that the initial information in the different channels is completely decoupled and there is no risk of information leakage. The quantitative analysis of crosstalk between different channels and the security analysis for the metakeys are discussed in Supplementary Notes 8 and 9, respectively. The results show that the proposed system has the capacity to resist to the interferences between 4 channels and maintaining channel independence.

The experimental fabrication and measurement setup of reported JEAM system are described in Fig. 6. Figure 6a is the schematic diagram of the experimental system and device, and Fig. 6b illustrates the experimental details of the reflected field. Figure 6c is a detail of a sample of JEAM, and Fig. 6d-e show photographs of transducer array to exhibit its fabrication details.

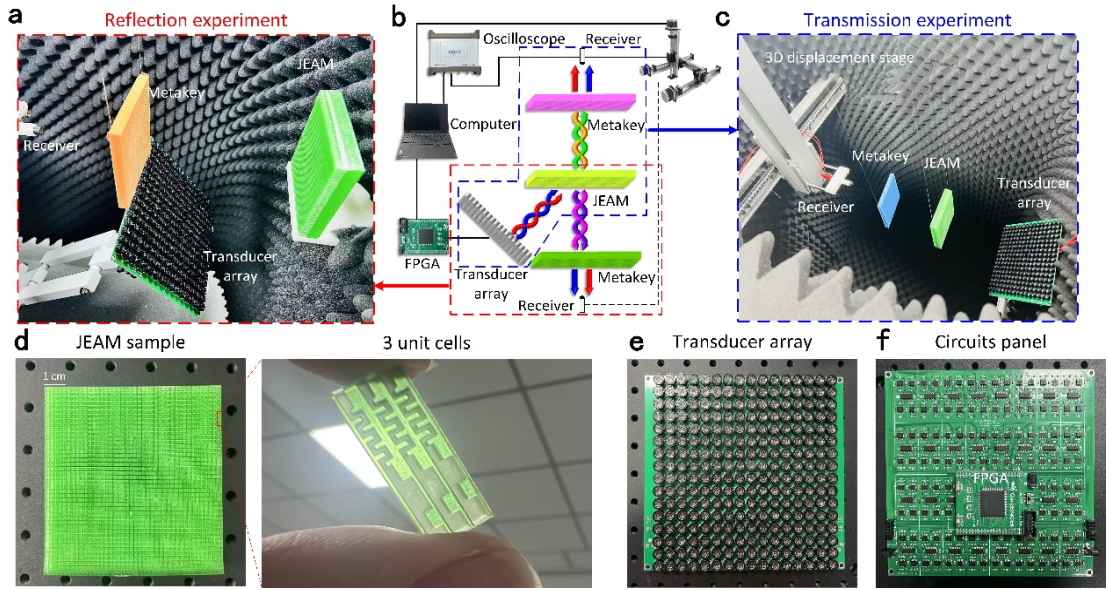

**Fig.6 The experimental equipment and samples. a** The photographs of reflection experiments. **b** The ultrasonic transducer array consists of 16×16 elements connected to the computer via an CoreEP4CE6 FPGA demo board. Ultrasonic receiver is mounted on a 3D translation stage, which is directly controlled by the computer through a MATLAB script. Received signals are captured by the oscilloscope and then transmitted to the computer. **c** The photographs of transmission experiments. **d** The photographs of the 3-D printing sample, consisting of 42 × 42 unit cells with a size of 9.86 cm × 9.86 cm × 1.8 cm. Scale bar, 1cm. **e**, **f** The photographs of the emitter arrays.

Figure 7 exhibits the acoustic intensity and phase distribution on the image plane of decrypted acoustic fields obtained by the numerical simulation and the experimental measurements, respectively. From the results, the predesigned letters of A, B for amplitude channel and C, D for phase channel are clearly observable, proving the high fidelity and confidentiality of the secure acoustic holography proposed here.

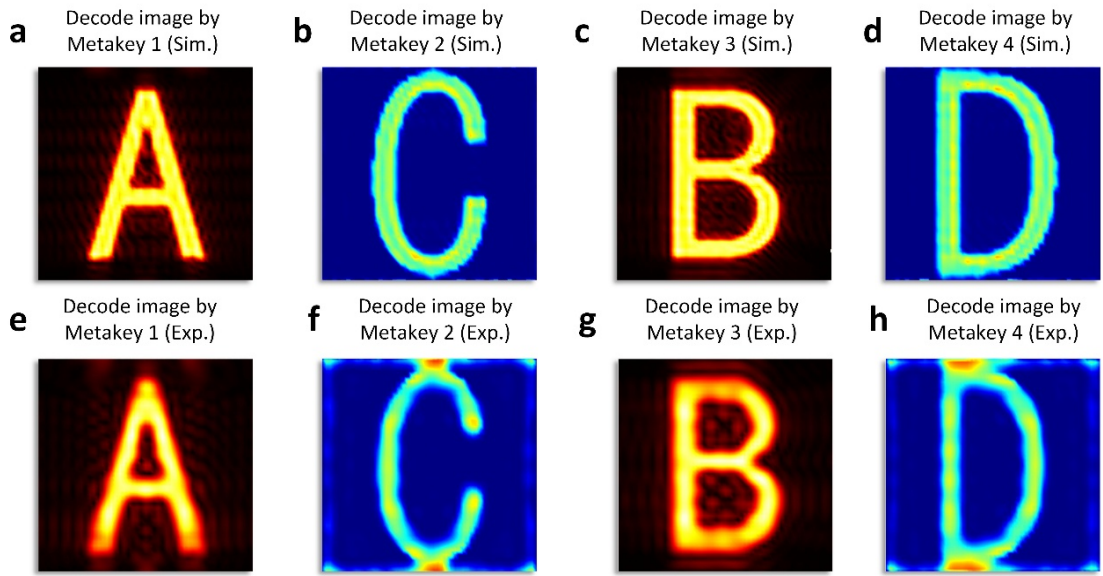

**Fig.7 The comparison between the results of experiments and simulations. a** Simulation results of the amplitude distribution in the sound field after transmission of sound waves from the transmitted encrypted field through the Metakey 1 at the operating position, experimental results **e**. **b** Simulation results of the phase distribution in the sound field after transmission of sound waves from the transmitted encrypted field through the Metakey 2 at the operating position, experimental results **f**. **c** Simulation results of the amplitude distribution in the sound field after transmission of sound waves from the reflected encrypted field through the Metakey 3 at the operating position, experimental results **g**. **d** Simulation results of the phase distribution in the sound field after transmission of sound waves from the reflected encrypted field through the Metakey 4 at the operating position,

experimental results **g**.

## Discussion

We have introduced and demonstrated the concept of Janus encrypted acoustic metascreen realizing two-faced asymmetric-encryption and decryption, which makes our system a single-input multiple-output, enabling subsequently a significant increase of the communication capacity and flexibility. Both the encoding and decoding of generated holograms are achieved by a metascreen that can separately manipulate the amplitude and the phase of the acoustic waves, and by different metastructures which constitute the public and private keys in the asymmetric encryption system. This flexible two-faced wavefront manipulation provides significant freedom to design various methods to encrypt sound waves. Furthermore, the phase hologram could also be extended to the field of optical encrypted holography, which allows the phase channel to transmit information independently even when the amplitude channel is disturbed. Regarding the two-faced communication achieved by the reported Janus metascreen, this concept could also inspire the establishment of an instant communication feedback mechanism, which could be widely used in the field of optical and electromagnetic communication.

We also have experimentally fabricated Janus encrypted metascreen and metakeys, and therefore, a series of Janus encrypted acoustic metascreen system have been numerically and experimentally demonstrated. Our finding constructs a new framework for encrypted acoustic communications with the advantages of multi-channel feature, two-faced manipulation, flexible encryption methods, and high fidelity for reconstructed image information. The results are beneficial for a variety of acoustic applications involving multi-channel acoustic information communications, complex noise control, acoustic illusion, etc.

On another hand, sound waves are more suitable for transmitting information in underwater environment or in other non-transparent media compared to optical and electromagnetic waves. Consequently, optical and electromagnetic encrypted communication in this media may no longer be the first option to consider. We have then the perspective to investigate the effect of fluid-solid coupling between metamaterials and fluids on acoustic waves, and to expand the operating bandwidth to MHz frequency range to make this encrypted communication method suitable for underwater environment. The influence (challenges, merits, and demerits, etc.) of fluid-solid coupling on acoustic manipulation and encryption in this work and underwater environment is discussed in Supplementary Note 10.

## Methods

**Numerical simulations**. The simulations are performed using the commercial finite element analysis software, COMSOL Multiphysics 6.2 with the module of Acoustic-Thermoviscous Acoustic Interaction, Frequency Domain". In simulations, the considered mass density and the sound speed of background medium air are $\rho$ = 1.21 kg/m³ and $c$ = 343 m/s, and solid materials in the unit cells are set as sound hard boundaries.

**Sample fabrications**. All the samples are made of UV Curable Resin and are manufactured via the 3D printing technique (Flight HT1001P, 0.1 mm in precision). The Young's modulus, Poisson's ratio, and density of the UV Curable Resin are $E_{CR}$ = 2.65GPa, $v_{CR}$ = 0.41, $\rho_{CR}$ = 1130kg/m³, respectively. The JEAM is consist of 42 × 42 unit cells with a size of 9.86 cm × 9.86 cm × 1.8 cm, and the metakeys are also consist of 42 × 42 unit cells with a size of 9.86 cm × 9.86 cm × 1 cm.

**Experimental measurements**. The transducer array consisting of 16 × 16 air-coupled ultrasonic transducers of 1 cm in diameter (Murata MA40S4S, driven by a 9-V peak-to-peak square-wave signal and producing a sinusoidal output at 40 kHz) are used as the source to generate signals. The proposed system can also be extended to other frequencies such as the audible range (Supplementary Note 11). The amplitude and phase emitted by each unit of the transducer array is controlled independently through an FPGA demo board[46]. Another air-coupled ultrasonic transducer (Murata

MA40S4R) is placed at 18 cm (For different operating distance refer to Supplementary Note 12) behind the sample of acoustic metakeys as a receiver to obtain the amplitude and phase of the sound field. The receiver is connected to a translation stage, which directly communicates with a laptop through a MATLAB script. Data from the receiver is collected via the oscilloscope (PicoScope 5444D) at the laptop for further operations. We used the receiver to scan 51×51 pixel points in an area of 10 cm×10 cm with the resolution of 1 mm, and then processed the data with the generation of holograms through MATLAB scripts, which took a total of 2 hours.

## Data availability

All relevant data that support the findings of this study are available from the corresponding authors upon reasonable request.

## Acknowledgments

This work was supported by the National Key Research and Development Program (No. 2022YFB3404300), the National Natural Science Foundation of China (No. 52272433), the Natural Science Foundation of Jiangsu Province (Grant No. BK20220798), Southeast University Interdisciplinary Research Program for Young Scholars (2024FGC1003), the Jiangsu Graduate Student Research Innovation Program (No. KYCX24_369), the SEU Innovation Capability Enhancement Plan for Doctoral Students (Grant No. CXJH_SEU 25037), and by IMPACT project LUE "I-META", part of the French PIA project "Lorraine Université d'Excellence" reference ANR-15-IDEX-04-LUE.

## Author contributions

B.A., Y.F.Z. and H.H.Z. proposed the concept. H.H.Z. performed the theoretical simulations. H.H.Z., Z.Y.H., X.G., X.H.H., Y.Y.M., T.T.L, F.L.M., H.Y.F., X.D.F., and W.W.K. designed and conducted the experiments. H.H.Z., T.X.Z., and G.D.Y. performed the mathematical studies. H.H.Z, Y.F.Z., H.Z., and B.A. wrote the manuscript. W.W.K., Y.F.Z., H.Z., and B.A. guided the research. All other authors contributed to data analysis and discussions.

## Competing interests

The authors declare that they have no competing interests.